\documentclass[twocolumn,iop,apj]{aastex63}
\usepackage{amsfonts,amsmath,graphicx,natbib,url,hyperref,nicefrac}
\usepackage{amssymb, verbatim, subfigure, paralist, soul, comment}
\usepackage{graphicx}
\usepackage{wrapfig}
\usepackage{verbatim}
\usepackage{lineno}
\usepackage{tikz}
\usepackage{color}

\usepackage{overpic}
\usepackage{multirow}
\usepackage{float}
\usepackage{CJKutf8}            
\usepackage[T1]{fontenc}
\usepackage{aecompl}
\usepackage{calligra}
\DeclareMathAlphabet{\mathcalligra}{T1}{calligra}{m}{n}
\DeclareFontShape{T1}{calligra}{m}{n}{<->s*[2.2]callig15}{}

\usepackage{ulem}
\usepackage{cancel}
\usepackage{hyperref}
\hypersetup{
    pdfnewwindow=true,      
    colorlinks=true,       
    linkcolor=blue,          
    citecolor=blue,        
    filecolor=blue,      
    urlcolor=blue           
}

\def\apjl{ApJL}               
\def\apjs{ApJS}               
\def\aap{A\&A}                

\DeclareMathAlphabet{\mathcalligra}{T1}{calligra}{m}{n}
\DeclareFontShape{T1}{calligra}{m}{n}{<->s*[2.2]callig15}{}

\hypersetup{colorlinks=true, urlcolor=blue, citecolor=blue}

\usepackage{comment}

\usepackage{color}
\usepackage{xcolor}

\newcommand{\OK}[1]{\textcolor{red}{[OK]}}

\newcommand{\pOK}[1]{\textcolor{green}{[OK]}}

\newcommand{\rlOK}[1]{\textcolor{orange}{[OK]}}

\shorttitle{} 
\shortauthors{Kirkeberg et al.}

\begin{document}

\title[]{On the Evolution of Disk-Embedded Binaries: Framing Local Models in Global Context}

\author[0009-0004-5111-6844]{Philip Kirkeberg}
\affiliation{Niels Bohr International Academy, Niels Bohr Institute, Blegdamsvej 17, DK-2100 Copenhagen Ø, Denmark}
\email{philip.kirkeberg@nbi.ku.dk}

\author[0000-0001-9222-4367]{Rixin Li
\begin{CJK*}{UTF8}{gkai}(李日新)\end{CJK*}}
\altaffiliation{51 Pegasi b Fellow}
\affiliation{Department of Astronomy, Theoretical Astrophysics Center, and Center for Integrative Planetary Science, University of California Berkeley, Berkeley, CA 94720-3411, USA}

\author[0000-0001-8716-3563]{Martin E. Pessah}
\affiliation{Niels Bohr International Academy, Niels Bohr Institute, Blegdamsvej 17, DK-2100 Copenhagen Ø, Denmark}

\begin{abstract}
The disks of Active Galactic Nuclei (AGN) have in recent years been recognized as possible sites for gravitational wave sources, leading to a series of numerical studies on the evolution of disk-embedded black hole binaries. The majority of these works have been carried out so far using the shearing box, a local Cartesian domain co-rotating with the binary center-of-mass around the supermassive black hole. The local nature of this framework allows for focusing computational power close to the binary at the expense of detaching the gas flow around the binary from the global dynamics.  In this paper, we provide a framework to assess the applicability of the shearing box for studying the long-term evolution of the orbital elements of the embedded binary in viscous hydrodynamic disks. We accomplish this by identifying the conditions under which relevant global timescales are longer than the gas-induced evolution timescale of the embedded binary across various AGN disk models. For black hole masses of interest, we report the existence of radii beyond which the global influence of the disk may be reasonably neglected, supporting the use of the shearing box. More generally, we introduce a systematic approach to link local simulations with the global problem they aim to approximate while providing a way to gauge their accuracy. This will prove to be essential as we seek to add additional physics, such as magnetic fields and radiative transport, to develop more realistic models for black hole binary mergers and their potential electromagnetic signatures in AGN disks.

\end{abstract}

\section{Introduction} 
Accretion disks surrounding supermassive black holes (SMBHs) in Active Galactic Nuclei (AGN) have recently emerged as promising sites for the formation and eventual merger of binary black holes (BBHs) ranging from stellar-mass to low-end intermediate-mass. These mergers produce gravitational waves such as those observed by the LIGO–Virgo–KAGRA Collaboration \citep{Ligo_2025}.
The inner regions of AGN are expected to be populated by stars and compact objects, either formed in-situ \citep{Stone_2017,Chen_2023} or captured from its surrounding nuclear star cluster \citep{Morris_1993,Fabj_2020}. This gives rise to a dense environment of compact objects prone to close encounters leading to gas-assisted binary formation \citep{DeLaurentiis_2023,Rowan_2023,Dodici_2024}; possibly enhanced by the presence of migration traps \citep{Bellovary_2016}.
Unless prompted to direct merger by further dynamical interactions \citep{Samsing_2022,Fabj_2024,Rowan_2025}, the subsequent evolution of a bound binary is governed by its interaction with the gaseous environment of the AGN disk. 

Although insights from isolated circumbinary systems \citep[e.g.][]{Munoz_2019,Tiede_2020,Dorazio_2021} can inform our understanding, the evolution of binaries embedded in AGN disks is fundamentally different. This setting introduces additional complexities: non-inertial forces in the binary frame, the tidal potential of the central SMBH, and the differential rotation and radial flows of the disk, which impose flow patterns on the circumbinary disk (CBD) that are neither axisymmetric nor geometrically thin.

This realization has led to a body of numerical simulations of disk-embedded BBHs. Early global studies by \citet{Baruteau_2011}, and more recently by \citet{Li_2021,Li_2022}, provided important initial insights but were limited to relatively massive binaries due to computational constraints. 
The computational challenge of running sufficiently high-resolution global simulations of realistic systems has in recent years been circumvented by the use of the shearing box approximation \citep{Hawley_1995}, which has enabled exploration of a wider part of parameter space, offering valuable insight into trends in the BBH evolution. 

Two-dimensional shearing-box studies by \citet{Li_Lai_2022} and subsequent extensions examined the effects of binary eccentricity, equation of state \citep{Li_Lai_2023}, and viscosity \citep{Li_Lai_2024}. Three-dimensional stratified shearing-box simulations followed, including studies of eccentricity \citep{Calcino_2024}, inclination \citep{Dittmann_2024}, and mass ratio \citep{Dittmann_2025}, building on the work of \citet{Dempsey_2022}. Most recently, shearing-box simulations by \citet{Mishra_2024} demonstrated that magnetohydrodynamic turbulence can produce circumsingle disks around individual black holes with strongly varying orientations, while \cite{Joshi_2025} demonstrated the possibility of collimated outflows due to build-up of magnetic pressure near the binary.

The shearing box has been a workhorse in astrophysical disk dynamics for decades, and it has the potential to provide valuable insights on disk-embedded binaries.  However, due to its local nature, the shearing box is agnostic to the global disk dynamics. Its local character makes it a useful framework because it allows to produce models representative of different disk radial locations spanning a wide range of physical conditions.  However, it is not immediately obvious how the local values needed to define a local model map into BBHs located at specific disk radii depending on the putative underlying global disk model.

In this work, we present a systematic way to link local BBH simulations to the global problem they aim to represent. 
This allows us to assess the applicability and robustness of the results obtained in the shearing box framework when addressing the long-term evolution of disk-embedded binaries. To this end, we provide a comprehensive overview of the most relevant parameters for the shearing box setup, an assessment of the validity of standard assumptions, and finally an evaluation of the physical scenarios in which typical shearing box simulations have the potential to accurately capture the evolution of BBHs in AGN disks. We accomplish this in the following fashion. In Section~\ref{sec: 2_embedded_bh} we introduce the relevant problem parameters, characteristic ratios and criteria for the validity of standard assumptions. Section~\ref{sec: localvglobal} presents the timescale criteria for assessing the validity of the local approximation when studying the long-term evolution of disk-embedded BBHs. In Section~\ref{sec: evaluating} we evaluate the parameters and criteria for realistic BBHs for a suite of AGN models. In Section~\ref{sec: discuss} we discuss our results, implications for future simulations of BBHs in AGN disks, and and some aspects of the problem we omit in our analysis. We highlight our most important findings in Section~\ref{sec: summary} and provide a brief guide on how to use our results in Appendix~\ref{sec:apply_framework}.

\section{The disk-embedded BBH}\label{sec: 2_embedded_bh}

We consider a circular BBH of total mass $m_{\rm b}$ embedded in the accretion disk of a SMBH with mass $M$. The binary semi-major axis is $a_{\rm b}$ and its center-of-mass is orbiting the SMBH on a circular orbit at a radial separation of $R_0$. The accretion disk is characterized by a radial surface density and scale height profile, $\Sigma(R)$, $H(R)$ with the local values at $R=R_0$ indicated by a subscript ``$0$''. 

We assume the dynamics of the system to be governed by the Euler equations coupled with the fluid Poisson equation and the binary's equations of motion.
The set of equations can be made dimensionless by defining four independent length-scale ratios, which, together with an equation of state and boundary conditions, uniquely determine the evolution of the system (modulo numerical parameters such as softening, mass-removal rate, etc.), and thus the evolution of the binary orbital parameters. 
In the following, we briefly introduce each of these ratios, their importance for the dynamics of the system, and the corresponding approximations as they apply to different asymptotic regimes.

\subsection{Characteristic ratios}
The following dimensionless ratios determine the binary evolution:
\begin{equation}\label{eq. problem_ratios}
    \frac{R_H}{a_b}, \:\: \frac{R_H}{H}, \:\: \frac{R_H}{R_0}, \:\: \frac{R_H}{\lambda_J}.
\end{equation}
Here,  $\lambda_{\rm J}=c_{\rm s}/\sqrt{G\rho}$ is the Jeans length and we have chosen as a reference scale the Hill radius
\begin{equation}
    R_{\rm H} = R_0\left(\frac{m_{\rm b}}{3M}\right)^{1/3}.
\end{equation}
These ratios encode information about \textit{binary compactness}, \textit{non-linearity}, \textit{degree of locality}, and \textit{self-gravity}, respectively. \\

\textit{Binary compactness} ---
The ratio between the Hill sphere and the semi-major axis of the binary describes its compactness relative to the tidal potential of the SMBH. For $a_b\gtrsim R_{\rm H}/3$ the binary system is unstable and prone to break-up \citep{Eggleton_1995}, while for $a_b<R_{\rm H}/3$ the binary is considered gravitationally bound and stable. 
Through dynamical binary formation \citep{Rowan_2023}, the initial binary semi-major axis is expected to be within one order of order magnitude of this limit, making the circumbinary systems significantly influenced by the surrounding differentially rotating disk environment.
In addition, although stable, binaries at these separations will present oscillations in orbital parameters, which may be important for the dynamics of the system \citep{Dempsey_2022}, see also Section~\ref{sec: caveats}. 
For tight binaries, i.e. $a_b\ll R_{\rm H}$, these oscillations diminish and the gas dynamics close to the binary is expected to exhibit features more similar to isolated circumbinary systems \citep[e.g.][]{Munoz_2019}, although this regime remains unexplored.

\textit{Non-linearity} --- 
Both the prominence of the background shear flow within the binary Hill sphere and the gravitational influence of the binary on the disk structure \citep{Ward_1997}, are set by the ratio between the binary Hill sphere and the disk scale height, $R_H/H$. This may be expressed in terms of the mass ratio $q=m_b/M$ and the disk aspect ratio $h=H/R$ or in terms of the binary mass to thermal mass ratio;
\begin{equation}\label{eq. R_H/H_or_q/h}
    3\left(\frac{R_H}{H}\right)^3=\frac{q}{h^3}=\frac{m_b}{m_{\rm th}}.
\end{equation}
These three expressions can be used interchangeably; we choose $q/h^3$ as the primary parameter. 

The thermal mass $m_{\rm th}=Mh^3$ is the limiting satellite mass for which the response of the disk is expected to be linear \citep{Goldreich_1980,Tanaka_2002} and above which the density waves launched by the satellite shock immediately leading to early gap formation. However, for $q/h^3<1$ small amplitude waves in inviscid disks will steepen and shock within some distance from the satellite \citep{Goodman_2001} initializing gap formation \citep{Zhu_2013,Cordwell_2024}.
The condition for gap formation depends on the efficiency of viscous stresses to refill the gap region \citep{Bryden_1999}. Demanding a gap to exhibit less than half the initial density, we state the viscous criteria as, see \citet{Kanagawa_2015},
\begin{equation}\label{eq. gap_formation}
    \frac{q}{\sqrt{25\alpha h^5}}\gtrsim1,
\end{equation}
which provides a proxy for when perturbing satellites embedded in a disk with angular momentum transport due to $\alpha$-viscosity \citep{Shakura_1973} are expected to form gaps. It should, however, be noted that a corresponding criterion in realistic inviscid magnetohydrodynamic disks with turbulence driven by the magnetorotational instability may be weaker \citep{Zhu_2013}. 

Based on select examples of the standard $\alpha$-viscosity disk models of \cite{Shakura_1973} and \cite{Sirko_2003} (SG03), \cite{Kaaz_2023} examined the potential of gap formation for disk-embedded binaries through the alternative criterion of \cite{Crida_2006}. They concluded that these systems rarely form gaps due to the low value of $q/h^3$ and relatively high viscosity. \cite{Dempsey_2022} and \cite{Dittmann_2024} performed a similar analysis, with single realizations of the models of SG03, \cite{Thompson_2005} (TQM05) and \cite{Dittmann_2020}, and generally found $q/h^3\lesssim 1$, except for the models of TQM05 which allowed $q/h^3\gtrsim 1$. We note that all of these studies explored a limited region of the AGN parameter space, considering similar values of SMBH mass, effective viscosities and SMBH accretion rates. 

\textit{Degree of locality} ---
Since $R_H/R\sim q^{1/3}$, the Hill radius of a typical BBH embedded in an AGN disk is much smaller than its radial separation from the SMBH. This justifies expanding the governing equations in orders of $R_H/R$ and thus the use a local framework.
We will discuss this further in Section~\ref{sec: localvglobal}. 

\textit{Self-gravity} --- 
The importance of self-gravity of the gas may on the one hand be gauged by the ratio between the Hill radius and the Jeans length. This determines whether a static homogenous sphere of radius $R_{\rm H}$ is prone to gravitational collapse. The flow within the Hill sphere is expected to deviate from the differentially rotating background, so that if $\lambda_{\rm J}\ll R_{\rm H}$, it may become gravitationally unstable even though the AGN disk itself is not inherently Toomre unstable.

On the other hand, we may compare the initial local mass within the Hill radius and the binary mass via
\begin{equation}\label{eq. disk_to_binary_mass_ratio}
    \frac{\Sigma_0R_H^2}{m_b} = 6\left(\frac{R_H}{\lambda_{\rm J}}\right)^2\left(\frac{R_H}{H}\right)^{-3},
\end{equation}
which instead informs about the relative gravitational influence of the local disk mass compared to the binary mass. We will refer to this mass ratio as the \textit{Hill-to-binary} mass ratio.
If the initial disk mass enclosed within the Hill radius satisfies
\begin{equation}\label{eq. disk_mass_limit}
    \frac{\Sigma_0R_H^2}{m_{\rm b}}>1
\end{equation} 
the binary does not strongly dominate the gravitational potential within the Hill sphere, signifying the importance of including self-gravity to accurately capture the gas dynamics.
\cite{Dempsey_2022} and \cite{Dittmann_2024} investigated typical values of $\Sigma_0R_H^2/{m_b}$ for BBHs in AGN disks. Both studies found most regions of the disks to have $\Sigma_0R_H^2/{m_b}\lesssim1$, thus supporting neglecting self-gravity. 
The Hill-to-binary mass ratio also gauges how strongly the gas back-reacts on the binary, thus determining the justifiability of keeping the binary orbital elements constant, as discussed in more detail below.

\subsection{Inspiral Timescale}\label{sec: inspiral_timescale}
Past studies of embedded BBHs generally aim to understand the gas-induced evolution of the binary orbital elements. 
To gauge the timescale of the semi-major axis evolution we introduce the inspiral timescale:
\begin{equation}
    \tau_{\rm inspiral} = \left\lvert\frac{\left<\dot{a}_b\right>}{a_b}\right\rvert^{-1},
\end{equation}
which is thus positive for both contracting and expanding binaries. The secular rate of change of the binary semi-major axis $\left<\dot{a}_b\right>$ is determined by the exchange of energy and angular momentum between the binary and the surrounding gas, through both gravitational interactions and accretion onto the BBH. 
If we consider $R_H$, $\Omega_0^{-1}$ and $\Sigma_0R_H^2$ as units of length, time and mass, the inspiral timescale may be written as 
\begin{equation}\label{eq. inspiral'}
    \tau_{\rm inspiral} =\frac{m_b}{\Sigma_0R_H^2}\Omega_0^{-1}\tau'_{\rm inspiral},
\end{equation}
where $\tau'_{\rm inspiral}$ is a dimensionless measure of the effectiveness of the surrounding gas at exchanging energy and angular momentum with the binary. In the isothermal and adiabatic limit, and neglecting self-gravity, $\tau'_{\rm inspiral}$ will be independent of the Hill-to-binary mass ratio $\Sigma_0R_{\rm H}^2/m_{\rm b}$, but will be non-trivially dependent on the three remaining independent ratios of Equation~\eqref{eq. problem_ratios}---in addition of course to complementary physics, boundary conditions and model parameters such as the sink radius and mass-removal rate. Once more realistic thermodynamics are introduced however, $\tau'_{\rm inspiral}$ may also depend on the Hill-to-binary mass ratio in a non-trivial way. In the local approximation $\tau'_{\rm inspiral}$ will additionally be independent of the ratio between the Hill radius and orbital radius $R_H/R_0$ leaving the compactness, $R_H/a_b$ and non-linearity, $q/h^3$, of the binary-disk system as the only independent ratios.

\begin{figure}[ht!]
\includegraphics[width=\linewidth]{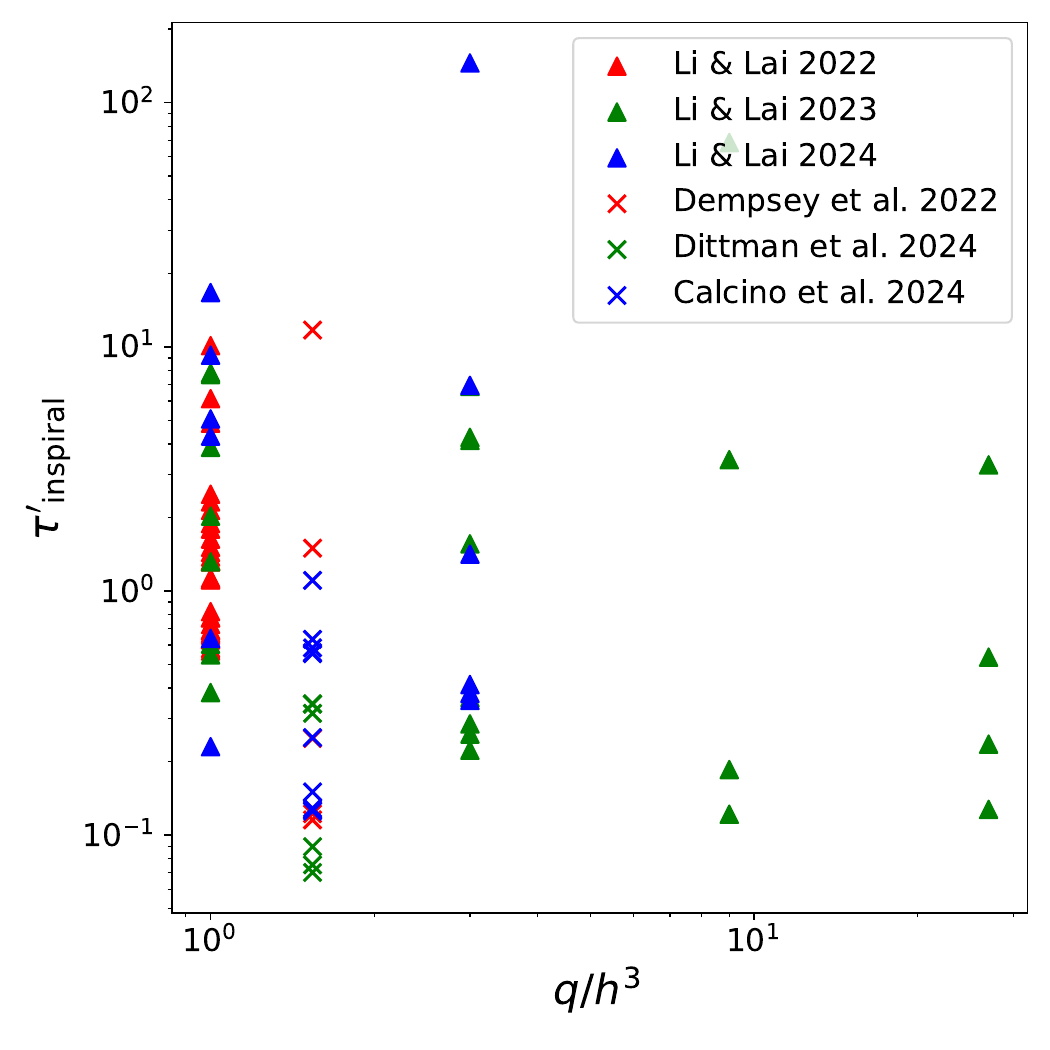}
\caption{The dimensionless measure of the inspiral timescale in Equation~\eqref{eq. inspiral'} measured in previous shearing box simulation, as a function of the chosen binary mass to thermal mass ratio $q/h^3$ in Equation~\eqref{eq. R_H/H_or_q/h}. 
\label{fig: sim_data_points}}
\end{figure}

In Figure~\ref{fig: sim_data_points} we show measurements of $\tau'_{\rm inspiral}$ from local simulations in the literature as a function of $q/h^3$. 
Although there is a significant variation arising from differences in binary compactness and the inclusion of various physical effects in the setups, it is generally observed that $\tau'_{\rm inspiral} \in \left[0.1,10\right]$. Our approach in this study is to remain agnostic about the intricate dependence of $\tau'_{\rm inspiral}$ on the problem parameters and state results in terms of $\tau'_{\rm inspiral}$. This implies that our conclusions will remain applicable as new insights on the values of $\tau'_{\rm inspiral}$ are obtained in the future from increasingly realistic simulations.

From Equation~\eqref{eq. inspiral'} it is apparent that greater local disk mass will lead to more rapid binary evolution due to stronger gravitational and accretion torques. This leads to an upper limit on the Hill-to-binary mass ratio, $\Sigma_0R_{\rm H}^2/m_{\rm b}<R_{\rm H}\tau'_{\rm inspiral}/a_{\rm b}$, beyond which the inspiral timescale is shorter than the binary orbital timescale. Beyond this limit it is therefore critical to self-consistently evolve the orbital parameters of the binary to accurately capture the binary evolution. We note that for $R_{\rm H}/a_{\rm b}>3, \: \tau'_{\rm inspiral}\sim1$, this limit leads to a weaker constraint on the disk mass than Equation~\eqref{eq. disk_mass_limit}. For the rest of this work we therefore use Equation~\eqref{eq. disk_mass_limit} as a proxy for both the importance of self-gravity and the live evolution of the binary orbital parameters.

\section{Local versus global}\label{sec: localvglobal}
The shearing box is a local approximation to the cylindrical disk equations, in which only terms of leading order in $c_{\rm s}/(\Omega_0R_0)$ are included \citep{Goldreich_1965,Hawley_1995}. This results in a framework in which the geometry around $R_0$ is Cartesian, as described in a coordinate system $(x,y,z)=(R-R_0,R_0\theta,z)$. The shearing box is thus deemed valid to the extent that $(x,y,z)/R_0\lesssim h$. For embedded BBH with $R_{\rm H}\sim H$, we would therefore expect the flow within the binary Hill sphere to be well represented in the shearing box formulation. 
Further, it has been found that the gravitational torque and power acting on the binary is strongly dominated by gas within the Hill sphere \citep[see e.g. Figure 5 of ][]{Li_Lai_2022}, in general suggesting that shearing box simulations of the evolution of embedded binaries are indeed representative of their `true' global evolution. 

However, the shearing box, by design, is agnostic of the radial flow of mass and angular momentum, and of the global reaction to the presence of the binary, in form of complete horseshoe orbits and potential gap formation. While the dynamics within the Hill sphere by definition is dominated by the binary potential, so that the gas morphology is expected to be largely independent of these effects, the supply of gas into the Hill sphere can be significantly influenced by these global effect.

In the following sections we identify three global timescales on which the global nature of the problem may influence the binary evolution. These are related to the horseshoe and radial flow of the disk and orbital evolution of the binary center-of-mass. 
By comparing them to the inspiral timescale of the binary, we gauge the validity of neglecting these effects, as it is effectively done in shearing box simulations.

\subsection{Horseshoe Flow}\label{sec: horseshoe_flow}
\begin{figure*}[ht!]
\includegraphics[width=1.02\linewidth]{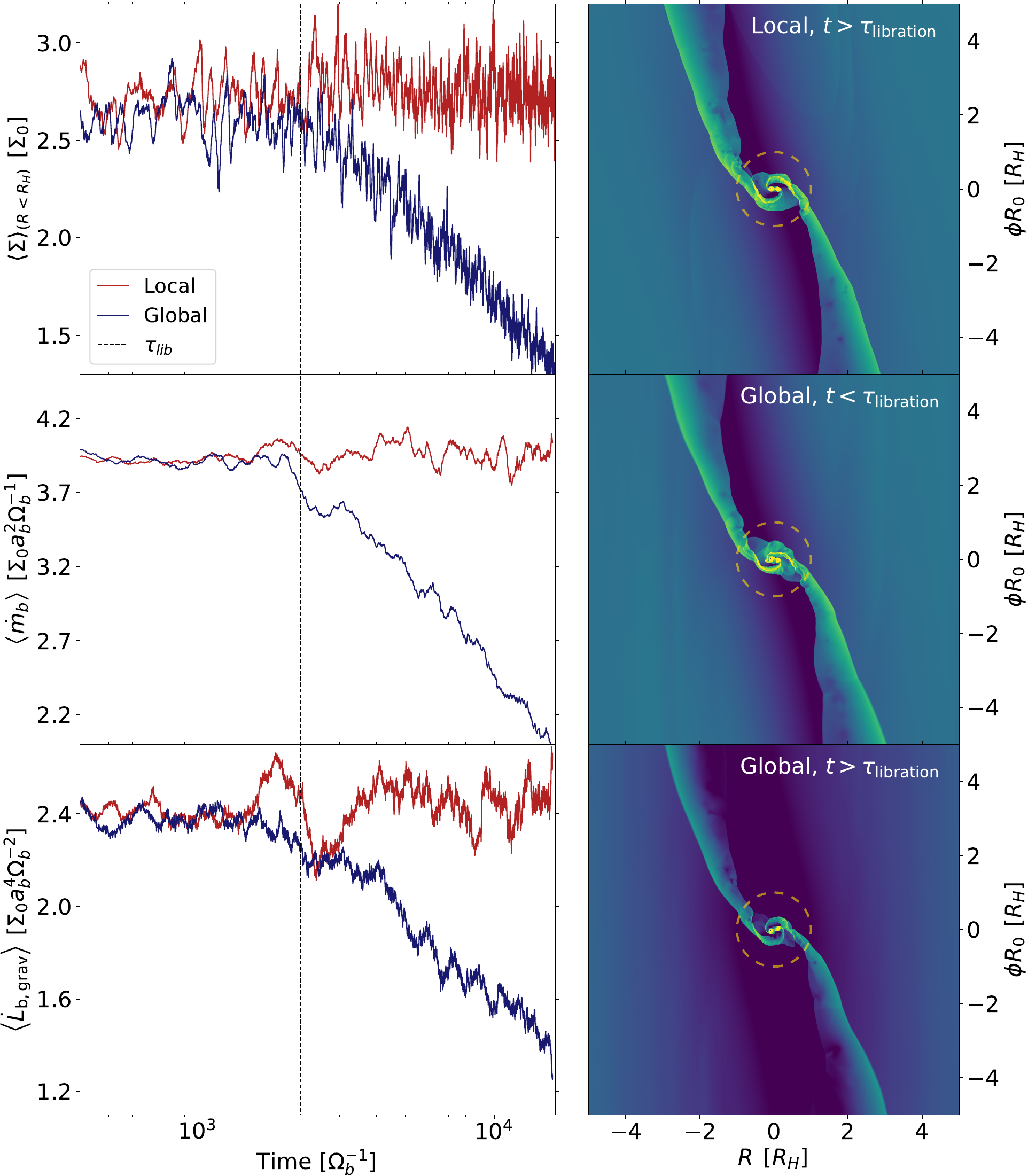}
\caption{Representative results from local and global hydrodynamical simulations of a disk-embedded binary. Left column: Average gas density within the Hill sphere (top) and rolling averages of binary accretion rate (middle) and gravitational torque (bottom) for local (red) and global (blue) simulations. Deviations appear after half a libration time as indicated by the dashed line. Right column: Density snapshot from local simulation before half a libration time (top), and from global simulations before (middle) and after (bottom). The dashed orange circle indicates the binary Hill sphere.}
\label{fig: numerical example}
\end{figure*}
Within the horseshoe region---a radially confined annulus centered on the SMBH and containing the binary---the gas follows horseshoe streamlines, executing U-turns near the binary and remaining nearly circular at greater distances \citep{Ward_1991}.
In this flow, an accreting embedded satellite is known to accrete material into its Hill sphere from streamlines located at the separatrix of its horseshoe region \citep{Lubow_1999}. Thus, to accurately capture the flow of mass and angular momentum into the binary Hill sphere, it is paramount to accurately capture the horseshoe flow and accretion streamlines on the relevant timescales.

\textit{Global response of horseshoe flow} --- Generally, three separate effects can influence the characteristics of the horseshoe flow, and all three have a characteristic timescale of the order of the libration timescale
\begin{equation}\label{eq. tau_libation}
    \tau_{\rm libration} = \frac{8\pi R_0}{3\Omega_0x_s},
\end{equation}
where $x_s$ is the half-width of the horseshoe region \citep{Jimenez_2017}
\begin{equation}\label{eq. shock_radius}
    x_s = H\frac{1.05\left(q/h^3\right)^{1/2}+3.4\left(q/h^3\right)^{7/3}}{1+2\left(q/h^3\right)^{2}}.
\end{equation}
This is the approximate time taken for a gas parcel located at $R_0(1+x_s)$ to complete two full orbits; i.e. a full horseshoe loop. Half of the libration timescale can thus be understood as the timescale for perturbations sourced locally around the binary, within the horseshoe region, to re-enter the local region through advection by the background flow.

\textit{i) Saturation of horseshoe flow}: 
low-mass planet migration studies suggest that the horseshoe flow can evolve through three regimes (characterized by linear, non-linear, or saturated non-linear corotation torques) depending on the planet's mass and the disk viscosity \citep[see, e.g.][]{Paardekooper_2009}.
While the evolution from linear to non-linear corotation torque is a local effect around the planet and happens within a few orbital periods, the saturation of the non-linear corotation torque has a characteristic time of the order of a few libration timescales, demanding continual strong interactions with the planet through several full horseshoe orbits to develop. 
The saturation of the horseshoe flow is thus a global response to the presence of the satellite, which may influence the accretion of mass and angular momentum into the binary Hill sphere through the decreased vortensity of the horseshoe region. 

\textit{ii) Gap formation:} as the binary mass increases beyond the thermal mass, it may potentially form a gap. Granted the viscosity is low enough, the gap will form on timescales of order the libration time. The gap will naturally deplete the density of the accretion-streamlines, thus lowering the accretion rate into the binary Hill sphere and influencing the net torque experienced by the binary. For sub-thermal binaries, the timescale of gap-formation through wave-steepening in inviscid disks is considerably longer, $\sim (q/h^3)^{-7/2}\,\tau_{\rm libration}$ \citep{Cordwell_2024}, and thus we will not consider it in this work.

\textit{iii) Accretion:} in a similar fashion, the simple fact that the binary is accreting material into its Hill sphere, and eventually onto the individual black holes, will lead to a depletion of the accretion streamlines---if they are not replenished by viscous or turbulent stresses.

We consider the combination of these effects as the global response to the presence of the binary, characterized by a timescale equal to half the libration timescale.
Critically, all three responses to the embedded binary depend on capturing complete horseshoe cycles. In shearing box simulations with damped boundary conditions \citep[e.g.][]{Li_Lai_2022,Li_Lai_2023,Li_Lai_2024}, the incoming streamlines are unaffected by the presence of the binary, thus making saturation of the horseshoe flow impossible due to the continual supply of vortensity, while also negating the feedback from accretion and wave shocking. For shearing box simulations with periodic azimuthal boundary conditions, the horseshoe orbits are indeed ``complete'', but of course the full dynamics is not captured due to the limited size of the domain. This leads to an effective libration timescale much shorter than the true libration timescale, which not only impacts the temporal evolution of the system, but possibly also the horseshoe flow itself. 
Indeed, hitherto all azimuthally periodic shearing box simulations are terminated before the density perturbations reach the azimuthal boundaries to avoid feedback, thus limiting the simulation time to less than the libration time  \citep[e.g.][]{Dempsey_2022,Dittmann_2024,Calcino_2024,Dittmann_2025}.

\textit{A concrete example} --- To illustrate the importance of the global response of the horseshoe flow, Figure~\ref{fig: numerical example} present results from hydrodynamical simulations performed using \texttt{Athena++} \citep{Stone_2020} to directly compare the evolution of an embedded binary modeled in a global disk and the same binary modeled in a shearing box. 

We consider a circular binary characterized by $q/h^3=10$ and $R_{\rm H}/a_{\rm b}=5$ embedded in an 2D inviscid and globally isothermal disk. For the global simulation we set $h=0.01$ at the binary center-of-mass, implying $q=10^{-5}$ and $R_0\simeq67R_{\rm H}$.
To mimic accretion onto the binary components, mass is removed at a rate $\Omega_{\rm b}$ from within $0.04a_{\rm b}$ of each component, through a torque-free sink prescription \citep[e.g.][]{Dittmann_2021}. 
In the local simulation the domain extends from $-50R_{\rm H}$ to $50R_{\rm H}$ in both $x$ and $y$, while the radial domain extends from $(R_0-33R_{\rm H})$ to $(R_0+67R_{\rm H})$ in the global simulation. To properly resolve the flow close to the binary, several levels of refinement (5 and 10 levels for the local and global simulations, respectively) are utilized to reach a resolution of $a_{\rm b}/\Delta x\sim140$ around the binary.
For the global simulation, we impose outflow conditions at the radial boundaries. For the local simulations, we employ relaxation boundary conditions in both radius and azimuth \citep[see e.g.][]{Li_2022}, so that the incoming flow resembles the unperturbed shearing flow.

Since the disk is inviscid and $q/h^3>1$, the binary is expected to form a gap. Therefore, from the discussion above, we expect a deviation between the two setups after half a libration time. The three left panels of Figure~\ref{fig: numerical example} show the time-evolution of the average density within the binary Hill sphere, the rolling average (over $10$ binary periods) of the accretion rate onto the binary, and finally the rolling average of the gravitational torque acting on the binary, for the local (red) and the global (blue) simulations. The vertical dashed line indicates half the libration timescale.
For $t<\tau_{\rm libration}/2$ the two simulations have reached an agreeing apparent quasi-steady state with constant density within the binary Hill sphere, binary accretion rate, and torque, indicating a balance between the accretion rate into the Hill sphere and onto the binary components. However, for $t>\tau_{\rm libration}/2$ the two start to deviate, with the density within the Hill sphere of the global simulation starting to decrease. This is due to the depletion of the accretion streamlines that results from the initiation of gap formation by wave-shocking and accretion. The reduced accretion into the Hill sphere disrupts the initial steady state, lowering the Hill density, which correspondingly leads to a decrease in binary accretion rate and gravitational torques. In the local simulation, this timescale does not exist as the incoming flow is, by construction, the initial shearing flow, unaffected by the response of the horseshoe flow to the presence of the binary. With no feedback from wave shocking or accretion, the accretion into the Hill sphere remains constant, and therefore the quasi-steady state remains immutable beyond the libration time. 

This example shows how the local framework is unable to capture the long-term evolution of the disk, possibly leading to a misrepresentation of the long-term evolution of the binary. 

\textit{Comparing to inspiral timescale} --- To gauge the importance of the global response of the horseshoe flow we introduce the ratio between half the libration timescale in Equation~\eqref{eq. tau_libation} and the inspiral timescale in Equation~\eqref{eq. inspiral'}
\begin{equation}\label{eq. tau_libation_ratio}
    \frac{\tau_{\rm {libration}}}{2\tau_{\rm inspiral}} = \frac{4\pi\Sigma_0 R_{\rm H}^2}{3m_b}\frac{R_0}{x_s}\tau'^{-1}_{\rm inspiral}. 
\end{equation}
For $\tau_{\rm {libration}}/(2\tau_{\rm inspiral})\gg1$ the global response happens on a longer timescale than the contraction or expansion of the binary, in which case the evolution of the binary may be well represented by results from local simulations. However, if $\tau_{\rm {libration}}/(2\tau_{\rm inspiral})\lesssim1$ the binary would not evolve significantly before half the libration time, making it necessary to perform global simulations in order to accurately capture the global response of the horseshoe flow to the embedded BBH. 

\subsection{Radial Flow}\label{sec: radial_flow}
A defining characteristic of the accretion disk is the inward radial flow of mass and corresponding outward transport of angular momentum, which is not captured in the local approximation. The flow of mass across the binary Hill sphere may however influence the accretion rate into the Hill sphere, and thus modify the steady average density surrounding the binary.

To gauge the importance of this flow we introduce the viscous timescale across the Hill radius
\begin{equation}\label{eq. tau_viscous}
    \tau_{\rm viscous} = \frac{R_H}{v_r},
\end{equation}
where $v_r$ is the radial velocity of the unperturbed disk. This is the timescale on which a total mass of $\Sigma_0R_H^2$, corresponding to the initial local disk mass, crosses the binary Hill sphere.
Comparing to the inspiral timescale we obtain
\begin{equation}\label{eq. tau_viscous_ratio}
\frac{\tau_{\rm viscous}}{\tau_{\rm inspiral}} = \frac{\Sigma_0R_H^3\Omega_0}{m_bv_r}\tau'^{-1}_{\rm inspiral},
\end{equation}
where $v_r$ is evaluated at the location of the binary center-of-mass.
For systems where $\tau_{\rm viscous}<\tau_{\rm inspiral}$, we may expect the global radial flow to modify the steady state torques acting on the binary.

If we assume angular momentum transport to be driven by an $\alpha$-viscosity \citep{Shakura_1973}, we can approximate the radial velocity as $v_r\sim\alpha c_{\rm s} h$. This allows for a direct comparison to half the libration timescale and thus a gauge of their relative importance:
\begin{align}\label{eq. viscous_libration}
\begin{split}
    \frac{2\tau_{\rm viscous}}{\tau_{\rm libration}} = & \frac{3^{2/3}}{4\pi\alpha}\left(\frac{q}{h^3}\right)^{1/3}\frac{x_s}{H},\\
\approx &\frac{3^{2/3}}{4\pi\alpha}\times
\begin{cases}
       1.05\left(q/h^3\right)^{5/6} &\quad\text{if } q/h^3\ll1\,,\\
       1.7\left(q/h^3\right)^{2/3} &\quad\text{if } q/h^3\gg1 \,.\\
\end{cases} 
\end{split}
\end{align}
For $\alpha\ll1$, the viscous timescale will therefore always be longer than the libration timescale for $q/h^3\ge 1$. However, for $q/h^3\ll 1$ the viscous timescale may become comparable or even shorter than the libration timescale, meaning that the radial flow across the binary could be equally relevant to the long-term binary evolution as the global response of the horseshoe flow.

\subsection{Orbital Evolution}\label{sec: orbital_evolution}
In addition to affecting the binary orbital elements, the binary-disk interaction will lead to a migration torque on the binary center-of-mass. However, due to the symmetric nature of the local approximation, this is not captured in shearing box simulations, and the center-of-mass is fixed at the center of the box (modulo a constant offset from a possible pressure gradient, see Section~\ref{sec: caveats}). For completeness, we investigate here the validity of this assumption for studying the binary evolution by comparing a typical migration timescale with the inspiral timescale.

For simplicity, we will assume the migration of the binary is well-approximated by Type I migration \citep{Goldreich_1980}, for which the center-of-mass torque has the general form
\begin{equation}
\Gamma = -K\Sigma_0\left(\frac{m_b}{M}\right)^2\Omega_0^2R_0^4h^{-2},
\end{equation}
where $K$ is a linear function of the density and temperature slopes of a locally isothermal disk. The exact value of $K$ is found from simulations and will depend on the viscosity and level of saturation, but in all cases will be of order unity \citep{Paardekooper_2009,Jimenez_2017,Tanaka_2024}. From the torque, we define the migration timescale $\tau_{\rm migration} = R_0/\dot{R}$, where $\dot{R}=2\Gamma R_0/j_p$ and $j_p=m_b\sqrt{GMR_0}$ is the orbital angular momentum of the binary with respect to the SMBH. The resulting migration timescale is then
\begin{equation}
\tau_{\rm migration} = \frac{1}{2\times3^{2/3}K}\frac{m_b}{\Sigma_0R_H^2}\frac{1}{h^2Q^{4/3}\Omega_0},
\end{equation}
which compared to the inspiral timescale is
\begin{equation}\label{eq. tau_migration_ratio}
\frac{\tau_{\rm migration}}{\tau_{\rm inspiral}} = \left[2\times3^{2/3}Kh^2\left(\frac{q}{h^3}\right)^{4/3}\tau'_{\rm inspiral}\right]^{-1}.
\end{equation}
Assuming $K,\tau'\sim1$ and  $h\lesssim 0.1$, this ratio is larger than one, except for when $q/h^3>1$---in which case the low mass assumption of Type I migration breaks down and the migration torque is expected to decrease significantly.
For a limiting Type I migration scenario of $h=0.1$ and $q/h^3=1$, we would need $\tau'_{\rm inspiral} \gtrsim 25$ in order for $\tau_{\rm inspiral}\gtrsim\tau_{\rm migration}$ which, while not unreasonable high, is not a typically measured value, as demonstrated in Figure~\ref{fig: sim_data_points}. 

We conclude that the orbital migration timescale of the binary center-of-mass for typical binaries embedded in AGN disks generally is longer than the inspiral timescale, making it unlikely to affect the binary orbital evolution.

\section{Evaluating timescales}\label{sec: evaluating}
Having identified the characteristic timescales on which the global nature of the problem may influence the binary evolution, we can now evaluate these for concrete examples of AGN-BBH systems. 
To this end, we need radial profiles of the disk scale height, surface density, and radial velocity. We obtain these by considering three widely used disk models; a single-zone standard $\alpha$-disk and the derivative models SG03 and TQM05. 

The standard $\alpha$-disk is constructed by parametrizing the viscosity through $\nu = \alpha c_{\rm s}H$ under the assumption of an axisymmetric, thin and optically thick disk \citep{Shakura_1973}. For specific opacity laws, this approach yields analytical solutions for the disk structure. However, these models are gravitationally unstable in the outer regions, as measured by the Toomre instability parameter
\begin{equation}
    Q = \frac{c_{\rm s}\Omega}{\pi G\Sigma}.
\end{equation}
The SG03 and TQM05 models extend the standard $\alpha$-disk to larger radii, aiming to explain the observed AGN luminosities by introducing additional pressure support to obtain solutions that are marginally stable to gravitational fragmentation. SG03 introduce a heating mechanism in the outer regions, which generates the radiation pressure necessary to keep $Q=1$. While the mechanism responsible for generating this heating is expected to be star formation, SG03 leave the necessary heating source unspecified. TQM05 extends upon this approach by modeling the feedback of star formation, leading to a solution in which mass advection is sourced by global torques instead of local viscous torques, and the accretion rate is no longer radially constant because of the mass loss due to star formation. 

\begin{figure}[ht!]
\includegraphics[width=\linewidth]{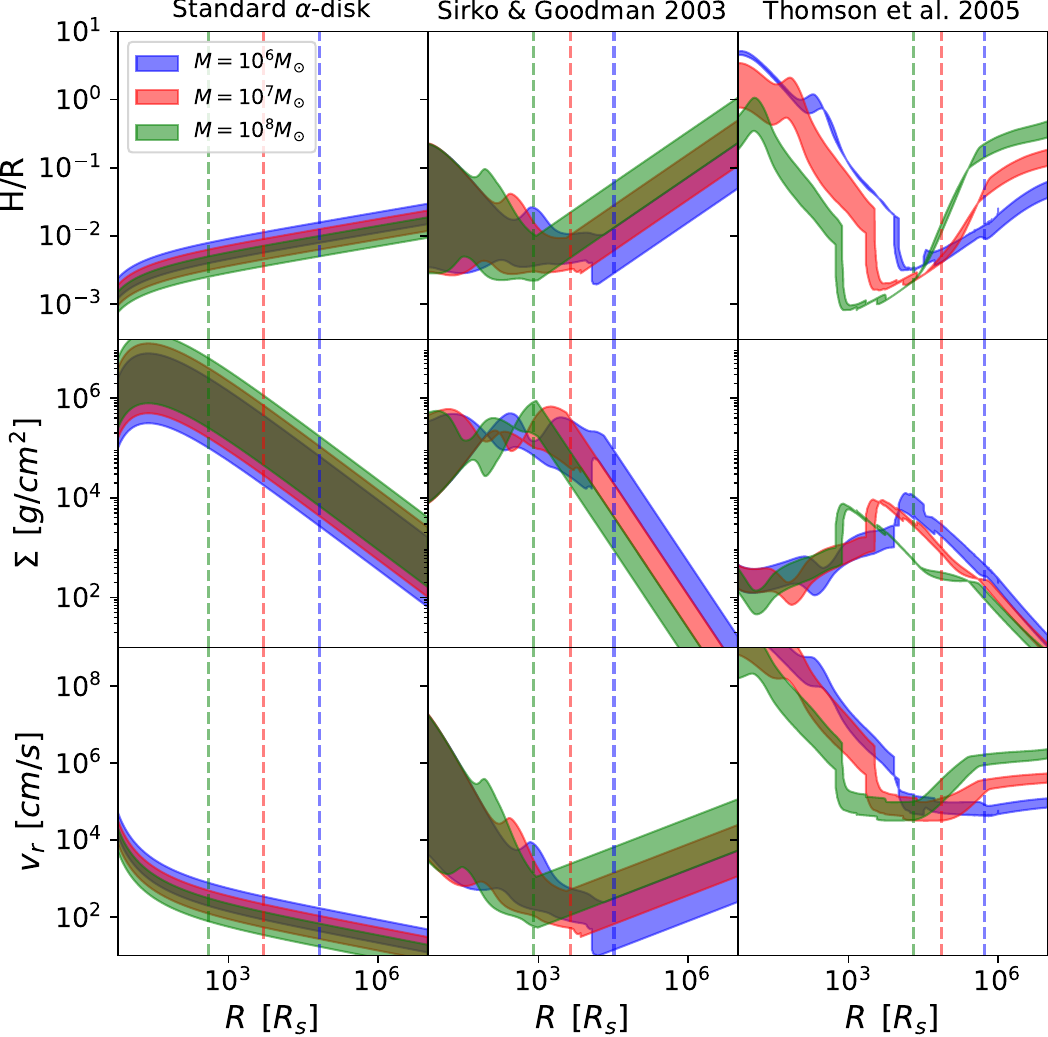}
\caption{Radial dependence of $H/R$, surface density $\Sigma$ and radial velocity $v_r$ for a standard $\alpha$-disk (left) and models of SQ03 \citep{Sirko_2003} (middle) and TQM05 \citep{Thompson_2005} (right). The models are shown for three different SMBH masses. For the $\alpha$-disk and SG03 model (both with $\alpha=0.01$) the shaded regions span $l_e\in [0.1,1]$ while for the TQM05 model they span $m\in[0.1,0.3]$.} 
\label{fig: agm_models}
\end{figure}

\subsection{Standard $\alpha$-disk}
We begin by presenting results for a single zone standard $\alpha$-disk, for which we assume the pressure to be dominated by gas pressure, and that the opacity is sourced by free-free absorption. 
Given a mass $M$ of the central SMBH, its accretion rate $\dot{M}$, and a constant value for the parameter $\alpha$, we obtain expressions for the radial profile of the surface density, scale height, mid-plane temperature, and radial velocity of the disk as
\begin{align}\label{eq. alpha_sigma}
    \begin{split} 
    \Sigma(R) = 4.6\times10^5 \alpha^{-4/5}l_e^{7/10}\left(\frac{M}{M_{\odot}}\right)^{1/5} \\ \left(\frac{R}{R_{\rm s}}\right)^{-3/4}f^{14/5} \:\rm g\:cm^{-2},
    \end{split}
\end{align}
\begin{align}\label{eq. alpha_h}
    \begin{split} 
    H(R) =   2.9\times10^3 \alpha^{-1/10}l_e^{3/20}\left(\frac{M} {M_{\odot}}\right)^{9/10} \\  \left(\frac{R}{R_{\rm s}}\right)^{9/8}f^{3/5} \:\rm cm,
    \end{split} 
\end{align}
\begin{align}\label{eq. alpha_T}
    \begin{split} 
    T_c(R) =   1.6\times10^8 \alpha^{-1/5}l_e^{3/10}\left(\frac{M} {M_{\odot}}\right)^{-1/5} \\  \left(\frac{R}{R_{\rm s}}\right)^{-3/4}f^{6/5} \:\rm K,
    \end{split} 
\end{align}
\begin{align}\label{eq. alpha_vr}
    \begin{split} 
    v_r(R) =   1.7\times10^6 \alpha^{4/5}l_e^{3/10}\left(\frac{M} {M_{\odot}}\right)^{-1/5} \\  \left(\frac{R}{R_{\rm s}}\right)^{-1/4}f^{-14/5} \:\rm cm \:s^{-1},
    \end{split} 
\end{align}
where $f=1-\sqrt{3R_{\rm s}/R}$ and $R_{\rm s}=2GM/c^2$ is the Schwarzschild radius of the central SMBH. Here the inner boundary of the disk to assumed to be located at the innermost stable orbit of the central SMBH. The Eddington fraction is $l_e=\dot{M}\eta c^2/L_{\rm edd}$ with Eddington luminosity $L_{\rm edd}=4\pi GMc/\kappa_{\rm es}$, and $\kappa_{\rm es}=0.2(1+X)\text{ cm}^{2}\text{g}^{-1}$ is the electron scattering opacity with electron fraction $X$. We assume $X=0.7$ and an accretion efficiency of $\eta=0.1$. 
In the left row of Figure~\ref{fig: agm_models} we show the profiles for $H/R$, $\Sigma$ and $v_r$ for $M\in\{10^6,10^7,10^8\}M_{\odot}$ and $\alpha=0.01$, with the color bands showing models with varying $l_e\in[0.01,1]$. As it is also clear from Equations~\eqref{eq. alpha_sigma} and \eqref{eq. alpha_h}, the $\alpha$-disk solutions are power laws for $R\gg R_{\rm s}$. The vertical dashed lines represent the radial distance from the SMBH beyond which the disk is Toomre unstable, for the respective SMBH masses, assuming $l_e=0.1$. 

\subsubsection{Timescale Ratios}
From Equations~\eqref{eq. alpha_sigma}-\eqref{eq. alpha_vr}, we express the timescale ratios defined by Equations~\eqref{eq. tau_libation_ratio}, \eqref{eq. tau_viscous_ratio}, and \eqref{eq. tau_migration_ratio} as:

\begin{align}\label{eq. alpha_libration_ratio}
    \begin{split}
        \frac{\tau_{\rm libration}}{2\tau_{\rm inspiral}}= 0.2\: f^{8/5}\left(\frac{\alpha}{0.01}\right)^{-3/5}\left(\frac{l_e}{0.1}\right)^{2/5}\\
        \left(\frac{M}{10^7M_{\odot}}\right)^{7/5}
        \left(\frac{R}{10^3R_{s}}\right)\\
        \left(\frac{\tau'_{\rm inspiral}}{10}\right)^{-1} \left(\frac{F(q/h^3)}{7}\right)^{-1},
    \end{split}
\end{align}
\begin{align}\label{eq. alpha_viscous_ratio}
    \begin{split}
        \frac{\tau_{\rm viscous}}{\tau_{\rm inspiral}} = 34\: f^{28/5} \left(\frac{\alpha}{0.01}\right)^{-8/5}\left(\frac{l_e}{0.1}\right)^{2/5}\\
        \left(\frac{M}{10^7M_{\odot}}\right)^{7/5}\left(\frac{R}{10^3R_{s}}\right) \left(\frac{\tau'_{\rm inspiral}}{10}\right)^{-1},
    \end{split}
\end{align}
\begin{align}\label{eq. alpha_migration_ratio}
    \begin{split}
        \frac{\tau_{\rm migration}}{\tau_{\rm inspiral}} = 10^{9}\:f^{-3/5} \left(\frac{\alpha}{0.01}\right)^{1/10}\left(\frac{l_e}{0.1}\right)^{-3/20}\\
        \left(\frac{M}{10^7M_{\odot}}\right)^{11/10}\left(\frac{R}{10^3R_{s}}\right)^{1/8} \\
        \left(\frac{\tau'_{\rm inspiral}}{10}\right)^{-1}\left(\frac{q/h^3}{10}\right)^{-4/3} ,
    \end{split}
\end{align}
where $F(q/h^3)=[1.05(q/h^3)^{5/6}+3.4(q/h^3)^{8/3}]/[1+2(q/h^3)^{2}]$ and $F(10)\sim 7$. 
These expressions suggest a difference in scale of the typical order of magnitude of each ratio. For the chosen parameters, half the libration timescale is of similar duration to the inspiral timescale, while the viscous and migration timescales are respectively two and nine orders of magnitude longer than the inspiral timescale. This suggest that both the radial flow and orbital migration may be safely neglected while studying the long-term binary evolution, but that the global response of the horseshoe flow could affect its evolution, for these representative values.

In addition, we may evaluate the Hill-to-binary mass ratio, given by Equation~\eqref{eq. disk_to_binary_mass_ratio}, as 
\begin{align}\label{eq. alpha_mass_ratio}
    \begin{split}
        \frac{\Sigma R_H^2}{m_b} = 0.01\:f^{11/5} \left(\frac{\alpha}{0.01}\right)^{-7/10}\left(\frac{l_e}{0.1}\right)^{11/20}\\
        \left(\frac{M}{10^7M_{\odot}}\right)^{13/10}\left(\frac{R}{10^3R_{s}}\right)^{9/8}\left(\frac{q/h^3}{10}\right)^{-1/3}.
    \end{split}
\end{align}
For the chosen values, the initial density within the Hill sphere is thus smaller than the binary, suggesting that keeping the binary fixed and neglecting self-gravity is justified.

\begin{figure*}[ht!]
\includegraphics[width=1.0\linewidth]{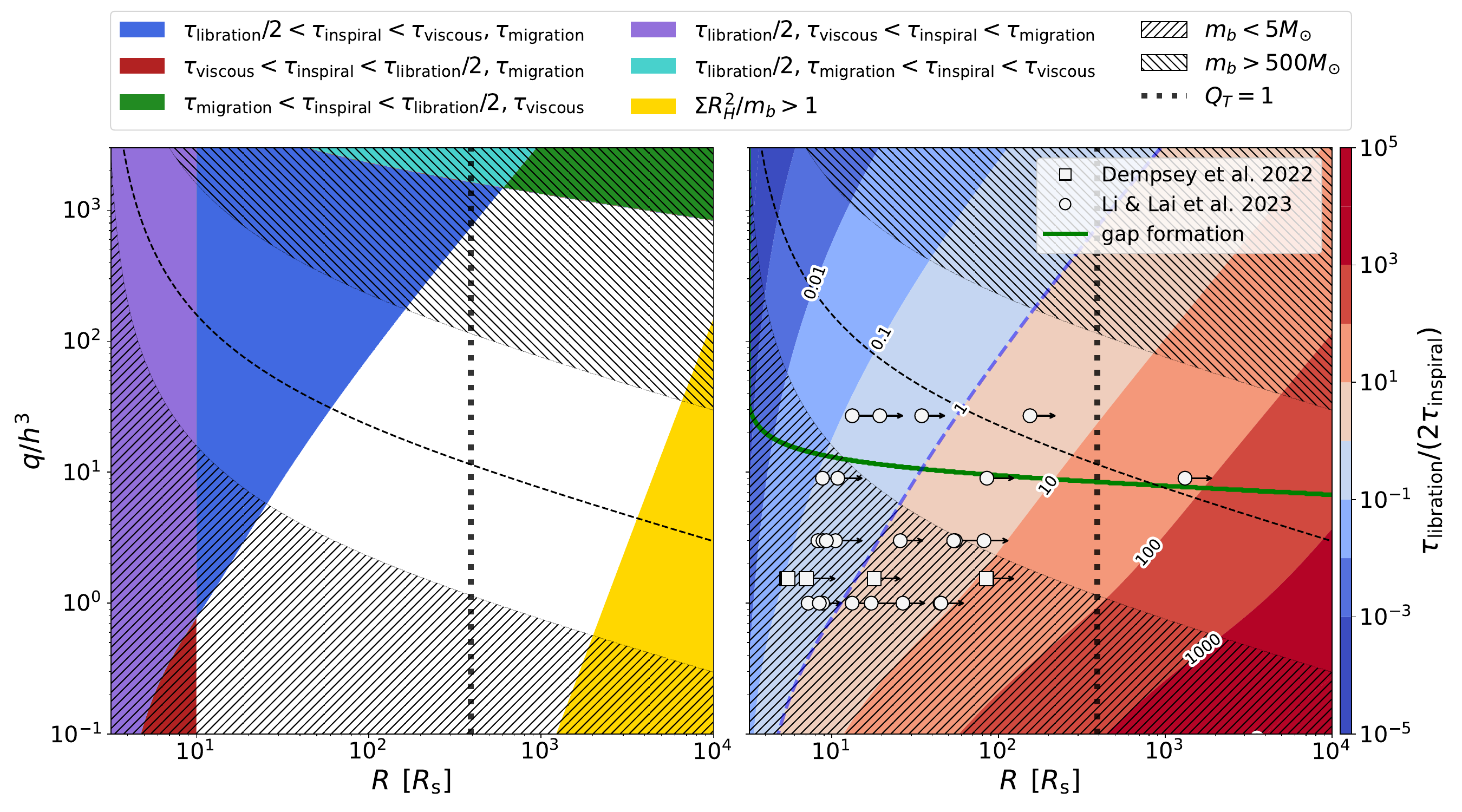}
\caption{Evaluation of timescale ratios in Equations \eqref{eq. tau_libation_ratio}, \eqref{eq. tau_viscous_ratio}, and \eqref{eq. tau_migration_ratio} for a standard $\alpha$-disk with SMBH mass $M=10^8M_{\odot}$, Eddington fraction $l_e=0.1$, and $\alpha=0.01$, assuming $\tau'_{\rm inspiral}=1$. The left- and right- tilted hashed regions have $m_b<5M_{\odot}$ and $m_b>500M_{\odot}$, respectively, while $Q\geq 1$ outwards of the bold dotted line.
Left panel: Coloured regions indicate parts of the $(R,q/h^3)$ parameter space for which the respective timescales are shorter than the inspiral timescale. In the yellow region the Hill mass is larger than the binary mass, see Equation~(\ref{eq. disk_mass_limit}).
Right panel: Contours of $\tau_{\rm libration}/(2\tau_{\rm inspiral})$. The squares and circles show the minimum radii at which the simulations presented in Figure~\ref{fig: sim_data_points} accurately capture the long-term binary evolution. Above the green line, the binary is expected to form a gap, as evaluated from Equation~\eqref{eq. gap_formation}.}
\label{fig: alpha_single}
\end{figure*}

\subsubsection{Regions of Parameter Space}
In the left panel of Figure~\ref{fig: alpha_single}, we visualize the relevant regions as given by Equations~\eqref{eq. alpha_libration_ratio}-\eqref{eq. alpha_mass_ratio} in the parameter space spanned by $q/h^3$ and $R$, for an $\alpha$-disk with $M=10^8M_{\odot}$, $l_e=0.1$ and $\alpha=0.01$. The right- and left-hashed regions represent parts of the parameter space for which $m_b<5M_{\odot}$ and $m_b>500M_{\odot}$, respectively, while the disk is Toomre unstable rightwards of the vertical dotted line. It is therefore only the non-hashed region that we deem the relevant part of parameter space for BBHs embedded in AGN disks, suggesting values of $q/h^3$ spanning from $\lesssim1$ in the outer part all the way to $\gtrsim10^3$ in the inner region. Additionally, since $m_{\rm b}\propto (q/h^3)M^{7/10}$, as seen from Equation~\eqref{eq. alpha_h}, the hashed regions will move upwards as $M$ decreases, implying the need for higher values of $q/h^3$ as the mass of the central object decreases.

The coloured patches indicate regions in which specific global timescales are shorter than the inspiral timescale, for $\tau'_{\rm inspiral}=1$. Most notable are the blue and purple regions in which $\tau_{\rm libration}/2<\tau_{\rm inspiral}$ spanning a range of radial distances and values of $q/h^3$. Within these regions, our analysis thus suggests that global simulations are necessary to accurately study the long-term binary evolution, since the binary will inspiral on a longer timescale than the global response of horseshoe orbits. 

The red and purple regions correspond to the parameter space where additionally the viscous timescale is shorter than the inspiral timescale. As anticipated in Section~\ref{sec: radial_flow}, the viscous timescale is shorter than half the libration timescale only when $q/h^3<1$. In these regions, the radial mass flow may influence the binary evolution.
In the yellow region, the local disk mass exceeds the binary mass, thus suggesting the importance of both live binary evolution and self-gravity of the gas. 
Finally, the green and cyan patches correspond to regions of parameter space where the Type I migration timescale is shorter than the inspiral timescale. As noted in Section~\ref{sec: orbital_evolution}, these are regions with $q/h^3\gg1$, in which the linear assumptions of Type I migration are not justified, and we therefore expect the binary to migrate significantly slower than inferred from the Type I migration torque. This indicates that the center-of-mass migration may safely be neglected when studying the evolution of the binary orbital parameters.

Expecting embedded binaries to mainly fall within the un-hashed region, Figure~\ref{fig: alpha_single} shows that both the global response of the horseshoe flow and self-gravity are important effects for this particular model, but that the evolution of binaries with $R_0\in[30,4000]R_{\rm s}$ is accurately captured by shearing box simulations neglecting self-gravity.

\subsubsection{Libration-to-Inspiral Timescale Ratio} 
In the right panel of Figure~\ref{fig: alpha_single}, we present contours of the libration-to-inspiral timescale ratio. For large distances $R$ from the central SMBH, the lines are broken power-laws as expected from Equations~\eqref{eq. viscous_libration} and \eqref{eq. alpha_libration_ratio}, increasing due to the increase of the binary Hill sphere. 

We assume $\tau'_{\rm inspiral}=1$, so that the contour of $\tau_{\rm libration}/(2\tau_{\rm inspiral})=1$ (blue dashed line) coincides with the limit of the blue region in the left panel. However, since the ratio is inversely proportional to $\tau'_{\rm inspiral}$, the limit of the blue region for the case with $\tau'_{\rm inspiral}=10$, would lie along the $\tau_{\rm libration}/(2\tau_{\rm inspiral})=10$ contour in the right panel. That is, for $\tau'_{\rm inspiral}=10$ we have $\tau_{\rm inspiral}>\tau_{\rm libration}/2$ for the whole region of $R\lesssim10^2R_{\rm s}$ and $m_b>5M_{\odot}$. The right panel may thus be used as follows: given a specific measurement of $\tau'_{\rm inspiral}$, we locate the contour equal to that value of $\tau'_{\rm inspiral}$ and then the region enclosed below this line is the part of the $(q/h^3,R)$ parameter space for which one may safely neglect global effects on the long-term evolution of the binary.

We also show selected simulation results from Figure~\ref{fig: sim_data_points} with the radial location of each data-point found by the above procedure. Each data-point represents the {\it minimum} radial separation from the SMBH where the inspiral timescale obtained in each simulation is shorter than half the libration timescale. The data points are generally bounded by the $\tau_{\rm libration}/(2\tau_{\rm libration})\in \{0.1,10\}$ contour lines as expected from Figure~\ref{fig: sim_data_points}, and most of the data points are located at small disk radii. This implies that it is justifiable to use local models for the long-term evolution of binaries throughout the majority of the disk. However, we also note that, for this particular disk model, the chosen values of $q/h^3<10$ correspond to low-mass binaries---except for the outermost region of the disk, where self-gravity of the gas within the Hill sphere starts to be important.
Finally, we also plot the gap formation criterion of Equation~\eqref{eq. gap_formation} as a green line. Although most of the binaries in the non-hashed region have $q/h^3\gtrsim1$ and thus facilitate gap formation, in the majority of the disk only BBHs with $m_b\gtrsim 50M_{\odot}$ satisfy the gap formation criterion.

It should be now evident that, for a given $\alpha$-disk model, figures like Figure~\ref{fig: alpha_single} can be used to determine under what conditions, i.e. values of $R/R_{\rm s}$, $q/h^3$, and $\tau'_{\rm inspiral}$, the shearing box can be safely used to evolve a disk-embedded binary. Similar considerations can be applied to more detailed disk models developed, as we discuss below.

\subsection{Global Disk Models -- SG03 \& TQM05}
In the middle and right panels of Figure~\ref{fig: agm_models}, we show the surface density, scale-height, and radial velocity of the SG03 and TQM05 disk models for $M\in \{10^6,10^7,10^8\}M_{\odot}$, computed with the \texttt{pAGN} package from \cite{Gangardt_2024}. Similar to the $\alpha$-disk, the shaded areas for SG03 represent models with $l_e\in[0.01,1]$, while for TQM05 they represent a variation in the angular momentum efficiency $m\in[0.1,0.3]$. The vertical dotted lines represent the transition from the inner gravitationally stable region to the outer marginally stable region of the disk with $Q\sim1$, shown for the respective model masses, with $l_e=0.1$ for SQ03 and $m=0.2$ for TQM05.
The SQ03 model resembles the $\alpha$-disk, with similar but slightly thicker and less massive inner and middle disk regions. The outer disk is hotter and so thicker than the $\alpha$-disk due to the auxiliary heating, which keeps it marginally stable. 
The TQM05 models have the typical bell-shape similar to SG03, but are, however, significantly thicker in the innermost disk region, while the middle region is thinner. In addition, the TQM05 model is considerably less massive than both the $\alpha$-disk and SG03 model, while its radial velocity is consequently faster, in order to maintain similar accretion rates. 
The SMBH mass dependence of the two are very similar; increasing the mass shifts the naturally thin stable disk region inwards, while keeping the general shape of the density and temperature profiles unchanged. For the TQM05 disk models, increasing the SMBH mass also leads to a decrease in the disk scale-height in the inner region and thickening of the outer disk.

\begin{figure*}[ht!]
\includegraphics[width=\linewidth]{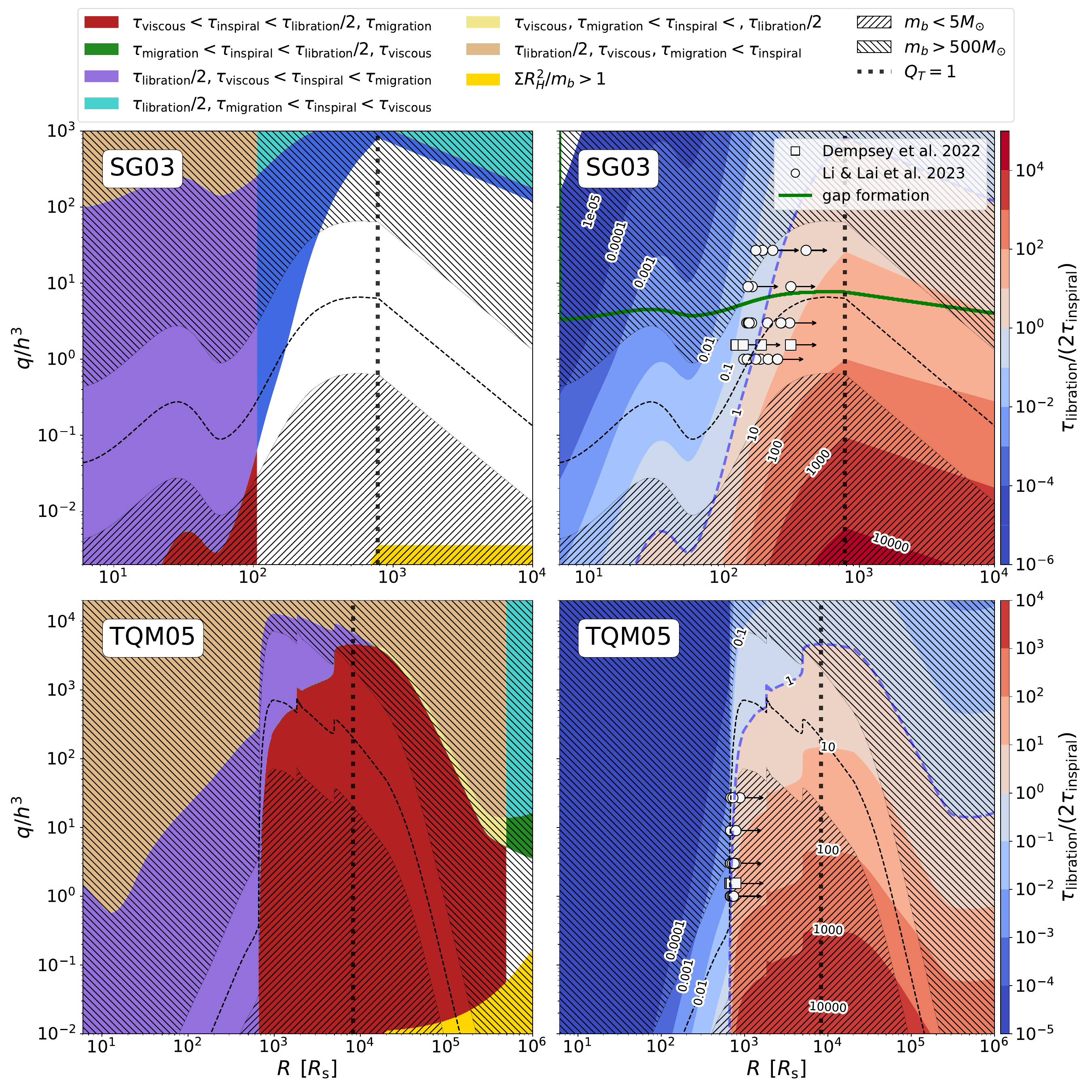}
\caption{Similar to Figure~\ref{fig: alpha_single} but for the SQ03 disk model with $l_e=0.1$, $\alpha=0.01$ (top) and the TQM05 disk model with $m=0.2$ (bottom), both with $M=10^8M_{\odot}$. Beyond the vertical dotted line the disk is marginally stable with Toomre's $Q\sim1$.}
\label{fig: sg_tho_single}
\end{figure*}

In Figure~\ref{fig: sg_tho_single} we show plots similar to Figure~\ref{fig: alpha_single} showcasing the ratios of Equations~\eqref{eq. disk_to_binary_mass_ratio}, \eqref{eq. tau_libation_ratio}, \eqref{eq. tau_viscous_ratio}, and \eqref{eq. tau_migration_ratio}, for SG03 (top row) with $l_e=0.1$ and $\alpha=0.01$ and TQM05 (lower row) with $m=0.2$---both with $M=10^8M_{\odot}$.

Aside from its multi-component structure, the contours of SG03 resemble those of the $\alpha$-disk shown in Figure~\ref{fig: alpha_single}. However, the generally thicker and less massive disk leads to a smaller $\tau_{\rm libration}/(2\tau_{\rm inspiral})$, stretching the blue and purple regions farther out. The thicker disk additionally leads to smaller values of $q/h^3$, which now spans from $\sim 0.4$ to $\sim60$ in the outer part of the gravitationally stable inner disk. This span encompasses the typical values chosen in past shearing box simulations, as seen in the top right panel.
The generally smaller $\tau_{\rm libration}/(2\tau_{\rm inspiral})$ pushes the data-points representing simulations farther out in the disk, making the results from these simulations only suitable for studying BBH located beyond $10^2R_{\rm s}$ in this particular disk model.
From the green line, we finally see that gap formation is only possible for large mass binaries in the thinnest region of the disk.

In the lower row of Figure~\ref{fig: sg_tho_single}, we show the corresponding results for a TQM05 disk model.
Due to the large radial variation of the disk scale-height, the allowed values of $q/h^3$ vary many orders of magnitude depending on the radial location of the binary, with a sharp increase at $R\sim 700R_{\rm s}$ owing to the nearly discontinuous decrease in $h$ as shown in Figure~\ref{fig: agm_models}. This gives rise to considerably larger values of $q/h^3$ in the middle region of the disk, now spanning from $\sim20$ to $\sim8000$, thus higher than those chosen in past studies (See Figure~\ref{fig: sim_data_points}). As a consequence of the even lower surface density of the TQM05 models, the regions with $\tau_{\rm inspiral}>\tau_{\rm libration}/2$ now extend out to $\sim10^3R_{\rm s}$. In addition, the regions in which $\tau_{\rm viscous}<\tau_{\rm inspiral}$ (shown with red, purple, and sand colors) now cover most of the parameter-space, due to the considerably faster radial velocity. This is an artifact of the global prescription of mass advection, where the radial velocity is not related to a local viscosity prescription, but set by a fraction of the sound speed, and so Equation~\eqref{eq. viscous_libration} no longer holds. For this TQM05 realization, the radial mass flow may thus be important for the evolution of BBH located all the way out to $R_0\sim5\times10^6R_{\rm s}$. Finally, similar to both the $\alpha$-disk and SG03 model, only in the outermost disk does the local disk mass exceed the binary mass.

\subsubsection{Variation of AGN Parameters}
In Figures~\ref{fig: sg_3x3} and \ref{fig: tho_3x3} we show plots similar to the panels on the right column of Figure~\ref{fig: sg_tho_single} for varying AGN parameters.

For SG03 (Figure~\ref{fig: sg_3x3}) the increase in SMBH mass and Eddington fraction leads to an increase in $h$ which decreases the expected values of $q/h^3$. This consequently leads to higher chance of gap-forming embedded BBHs for AGNs with the lowest SMBH masses and Eddington fractions. The expected values of $q/h^3$ increase with $\alpha$, however, due to the dependence of Equation~\eqref{eq. gap_formation} on the value of $\alpha$ this only moderately affects the fraction of gap-forming BBHs. The surface density generally decreases with increasing SMBH mass and $\alpha$, and decreasing Eddington fraction, leading to a larger $\tau_{\rm libration}/(2\tau_{\rm inpsiral})$ fraction and thus shifting the contours towards smaller radii. 
This means that shearing box simulations are expected to accurately capture the binary evolution within a shorter distance from the SMBH for AGN disk with larger SMBH masses, accreting at lower rates, and with higher effective disk viscosities.
However, since the transition from a stable to marginally stable disk also changes with SMBH mass and Eddington fraction, it is still only in the outermost part of the inner stable disk that the shearing box simulations are deemed appropriate. For all disks, the local disk mass is consistently lower than the binary mass.

Figure~\ref{fig: tho_3x3} shows similar plots for the TQM05 models. Compared to SG03, these models exhibit less obvious variation with SMBH mass and accretion efficiency. However, we note that the radial location of the sharp decrease in the disk aspect ratio moves inward as the SMBH mass increases, which shifts the contours inwards and upwards thus allowing smaller radii of the disk to be accurately studied by local simulations. 

\section{Discussion}\label{sec: discuss}
By comparing the binary inspiral timescale with other relevant global disk dynamics timescales, we have assessed the potential of local simulations to accurately model the evolution of embedded BBHs in AGN disks. In the following, we discuss some key issues related to the global disk response and radial gas flow, highlighting potential caveats of our approach.

\subsection{Importance of Global Disk Response}
Based on the comparison between $\tau_{\rm lib}/2$ and $\tau_{\rm inspiral}$, our analysis suggest that there exist a minimum AGN radius---ranging from $10^2R_{\rm s}$ to $10^5R_{\rm s}$, depending on the AGN disk model, for a typical $50M_\odot$ binary--- beyond which the global response to the presence of the binary can be safely neglected. 
However, the statement that this is accurately modeled by the ratio $\tau_{\rm lib}/(2\tau_{\rm inspiral})$ deserves additional discussion.

\textit{Strength of global response} --- In parts of the parameter-space where $\tau_{\rm lib}/(2\tau_{\rm inspiral})<1$, the global response to the presence of the binary \textit{can} propagate to the local region of the binary faster than the binary evolves. This does however not necessarily mean that it does, or that the effect on the binary evolution will be significant. 
For turbulent flows---or those with a high effective viscosity---the accretion streamlines may be sufficiently replenished, re-saturating the co-rotation region and effectively washing out the global response of the disk. 
Similarly systems with $q/h^3\ll1$, i.e. with binary mass much smaller than the thermal mass, are expected to hardly disturb the disk structure, or do so on a significantly longer timescale \citep{Cordwell_2024}, diminishing the effect of the global response to its presence. These effects may be partly captured by additionally considering some gap-formation criterion, like Equation~\eqref{eq. gap_formation}. From Figure~\ref{fig: sg_3x3} it is apparent that this significantly extends the region of applicability of the local approximation. 

The numerical example shown in Figure~\ref{fig: numerical example} should only be considered a demonstration of the existence and related timescale of the global response, and not a gauge for the strength of its effect on the binary evolution---which remains unexplored. The strength of this effect is expected to be affected by e.g. magnetohydrodynamic turbulence or viscosity, thermodynamics, and the extension to 3D flows.

\textit{Hill density modeling} --- It is expected that the global response of the disk mainly affects the average density, and not the morphology, of the gas within the binary Hill sphere. This suggests that it may be possible to separately model the \textit{global} average Hill density, $\left<\Sigma\right>^{\rm G}_{<R_H}$, and combine this with detailed knowledge of the local dynamics as obtained from local simulations. For example, by measuring both $\tau'_{\rm inspiral}$ and the \textit{local} average Hill density $\left<\Sigma\right>^{\rm L}_{<R_H}$, from local simulations, an estimate of the inspiral timescale including the effect of the global response would be
\begin{equation}
    \tau_{\rm inspiral}=\Sigma_0\,\frac{\left<\Sigma\right>^{\rm G}_{<R_H}}{\left<\Sigma\right>^{\rm L}_{<R_H}}\,\tau'_{\rm inspiral}.
\end{equation}
To first order the global average Hill density may be inferred from the gap density as found for massive planets \citep[e.g.][]{Duffell2020}, through $\left<\Sigma\right>^{\rm G}_{r<R_H}=(\Sigma_{\rm gap}/\Sigma_0)\left<\Sigma\right>^{\rm L}_{<R_H}$. In this way, results from local simulations could be applied to all regions of the AGN disk, independently of the timescale of the global response.

\subsection{Importance of Radial Flow}
Our results on the importance of the radial flow across the Hill sphere warrant some discussion.  We have chosen to gauge the potential impact of the radial flow through the viscous timescale across the Hill sphere (Equation~\eqref{eq. tau_viscous}). That is, we suggest that the radial flow may impact the binary evolution if its time for replenishing the Hill sphere is shorter than the inspiral timescale. While the corresponding timescale for the global response of the binary (Equation~\eqref{eq. tau_libation}) signifies a hard temporal transition before which we expect no local impact from this effect, the radial flow instead presents a continuous effect on the binary, of generally unknown strength. 
We therefore remain aware of the limitations of our results regarding the importance of radial flow—especially for the models of TQM05—and suggest careful assessments when applying shearing box results in these cases.

\subsection{Other Relevant Processes}\label{sec: caveats}
When comparing the local and global disk models above, we have omitted some aspects that may impact the assessment of the appropriateness of the shearing box in capturing the relevant gas dynamics accurately. We briefly discuss and assess these issues below.

\textit{3-body dynamics} --- A binary in the tidal potential of a central SMBH is stable if it is well-contained within its Hill sphere. However, its semi-major axis, eccentricity, and inclination may oscillate in time. These effects can nevertheless be included in the local framework by either integrating the appropriate Hills equations \citep[e.g.][]{Binney_2008} or the full equations of motion of the binary under the influence of the tidal potential, as is done in e.g. \cite{Dempsey_2022}, or prescribing the precession in an approximate manner where appropriate in 2D \citep[e.g.][]{Li_Lai_2022}. We therefore do not consider this a limitation to the applicability of local models for studying the evolution of embedded BBHs.

\textit{Non-circular outer orbits} --- In our analysis, we have assumed the orbit of the binary center-of-mass around the SMBH to be circular. 
For non-zero eccentricity and inclination less than the half-width of the horseshoe region, as defined in Equation~\eqref{eq. shock_radius}, the libration timescale will be similar to the circular case, and our results and framework remain unaltered. While our analysis would be invalid for systems with larger eccentricity and inclination, we note that deviations from circular orbits on order the disk scale height are in general not well captured in local simulations. Our framework, therefore, remains valid for those eccentric and inclined systems with $q/h^3\gtrsim 0.5$ for which local simulations are a priori an appropriate alternative to global simulations.

\textit{Non-circular binary orbits} --- Our framework focuses on the timescale for the evolution of the semi-major axis of the binary while neglecting changes in its eccentricity or inclination. Our approach, however, could easily accommodate for these and other binary timescales through the dimensional inspiral timescale $\tau'_{\rm inspiral}$ in  Equation~(\ref{eq. tau_viscous_ratio}). We note that the eccentricity and inclination of disk-embedded binaries have been found to evolve faster than their semi-major axis \citep{Calcino_2024, Dittmann_2024}. This hierarchy positively impacts the applicability of the local model by moving inwards the blue dashed line in Figure \ref{fig: sg_3x3} and \ref{fig: tho_3x3}, described in Appendix~\ref{sec:apply_framework}.

\textit{Indirect source terms} --- In most numerical work on disk-embedded bodies, the coordinate frame is anchored on the central massive object rather than to the system’s barycenter. This choice introduces non-inertial forces, referred to as indirect forces.
Including these terms in the governing disk equations is crucial to correctly capture the global gas dynamics, for example the appearance of all five Lagrange points in the case of a single embedded satellite \citep{Crida_2025}. 
The standard shearing box framework naturally includes the rotational non-inertial terms due to its motion around the central object.
However, local simulations of embedded objects often omit the \textit{indirect} terms, assuming they are negligible compared with the \textit{direct} contributions from the central mass and the embedded bodies.
 To assess the relative strength of these source terms in the local framework, we consider the case of a single embedded satellite of mass $qM$ located at $\mathbf{R}_0=(R_0,\theta_0,z_0)$ with respect to a central object of mass $M$ on which the frame is fixed. In this case, the corresponding source term will be the gradient of the sum of the three potentials
\begin{align}
    \Phi_{\rm C}(\mathbf{R}) &= -\frac{GM}{\lvert\mathbf{R}\lvert}, \hspace{5pt} \Phi_{\rm D}(\mathbf{R}) =-\frac{qGM}{\lvert \mathbf{R}_0-\mathbf{R}\lvert},\\     
    &\Phi_{\rm I}(\mathbf{R}) =-\frac{qGM}{R_0^3}\mathbf{R}_0\cdot\mathbf{R},
\end{align}
where the $\Phi_{\rm C}$ and $\Phi_{\rm D}$ are the gravitational potentials of the central object and the satellite, respectively, and $\Phi_{\rm I}$ is the indirect term arising from the non-inertial nature of the coordinate frame. Introducing local coordinates $(x,y,z)=\left(R-R_0,R_0(\theta-\theta_0),z-z_0\right)$ and assuming $x/R_0,y/R_0,z/R_0 \sim h\ll 1$, the strength of the corresponding source terms, $S_i=\nabla\Phi_i$, can be approximated as
\begin{equation}
    S_{\rm C}\sim\Omega_0^2R_0, \hspace{10pt} S_{\rm D}\sim q\Omega_0^2R_0h^{-2}, \hspace{10pt} S_{\rm I}\sim q\Omega_0^2R_0,
\end{equation}
For $q\ll 1$ the indirect term is subdominant with respect to the central source term, while smaller than the direct satellite source term for $h\ll 1$, as it is assumed in the local approximation. Given that $q\ll1$ for typical systems of embedded BBHs, we conclude that the indirect source terms can safely be omitted in the local framework considered here.

\textit{Initial disk structure} --- In our analysis we assume an unperturbed disk structure prior to the onset of the binary evolution. While substructures present in the disk caused by its presumably inherent populations of satellites may influence the binary evolution, we consider this an issue of local and global simulations alike, and thus not a significant caveat of our study on the applicability of the local approximation.
It nevertheless remains an important concern, which demands serious modeling to understand the proper initial conditions for BBH simulations, or simulations exploring binary-formation and subsequent evolution self-consistently.

\textit{Disk-wind} --- The disk orbital frequency deviates from its Keplerian value in the presence of a radial pressure gradient.
Expanded to second order in $h$, and neglecting vertical stratification, the disk orbital frequency is \citep[e.g.][]{Tanaka_2002}
\begin{equation}
    \Omega \simeq \Omega_0\left(1-\frac{1}{2}h^2\beta\right),
\end{equation}
where we have assumed a radial power-law dependence for the pressure $p\propto r^{-\beta}$. 
The deviation from the Keplerian frequency at $R_0$ thus creates a headwind (or tailwind, depending on the sign of the pressure gradient) experienced by the binary orbiting with angular frequency $\Omega_0$. Similarly to the radial flow in Equation~\eqref{eq. tau_viscous} we can introduce a headwind timescale
\begin{equation}\label{eq: tau_headwind}
    \tau_{\rm headwind} = \frac{R_H}{R_0(\Omega_0-\Omega)} \simeq \frac{2R_{\rm H}}{Rh^2\Omega_0\beta} \,,
\end{equation}
which gauges the time taken for a total mass of $\Sigma R_{\rm H}^2$ to cross the binary Hill sphere due to the sub or super-Keplerian frequency of the disk.
Comparing Equation~\eqref{eq: tau_headwind} directly to the viscous timescale, again assuming $v_r\sim\alpha c_SH$, we find
\begin{equation}
    \frac{\tau_{\rm headwind}}{\tau_{\rm viscous}} = \frac{2\alpha}{|\beta|}.
\end{equation}
For $\alpha\ll1$ and $\beta\sim1$, the timescale related to the disk-wind is therefore significantly shorter than that of the radial flow, indicating the relative importance of including the effect of a global pressure gradient. However, in contrast to the radial flow, the disk-wind experienced by the embedded binary can be included in shearing box simulations, assuming a constant radial pressure gradient across the box. This can be achieved by radially displacing the binary center-of-mass and the point at which the gas corotates with the box. For this reason, we have omitted this effect in our main analysis. We do however note that since the timescale decreases for thicker disks, at fixed $q/h^3$---similarly to the viscous timescale---the region of the disk in which the radial flow is important, would similarly be impacted by the differential disk-wind.

\textit{Additional physics} --- 
Finally we note that our conclusions are largely unaffected by the complexity of the disk physics included in the simulations. Even though the example in Figure~\ref{fig: numerical example} is deliberately simple, the characteristic timescale of the global response would remain unchanged upon introducing additional physics, and any variations in the binary evolution would be absorbed into the dimensionless parameter $\tau'_{\rm inspiral}$ (Equation~\eqref{eq. inspiral'}). The main caveat is that $\tau'_{\rm inspiral}$ is assumed to be independent of the Hill-to-binary mass ratio, an assumption tied to the simplified thermodynamics. While more sophisticated thermodynamics would break this independence, we expect any resulting correction to be relatively modest and not qualitatively change our conclusions.

\section{Summary}\label{sec: summary}
The shearing box has proved to be a valuable tool for the study of BBHs in AGN disks and helped advance our understanding of their long-term evolution. However, the interpretation and validity of these results may be affected by the inherent neglect of global disk dynamics. In this work we have quantified this by computing timescale ratios gauging the importance of the global response of the disk to the presence of the binary, the radial flow and the orbital migration of the binary. 

\subsection{Some General Lessons}
Considering typical values of interest for BBHs embedded in AGN disks our main conclusions can be illustrated as follows:
\begin{itemize}
    \item For binaries located in the middle thin disk region of the AGN, $q/h^3$ can vary between $\sim[1, 10^3]$ for the disk models of SG03 and $\sim[10^2, 10^4]$ for those of TQM05. This motivates exploring much larger values of $q/h^3$ than done hitherto (see Figure~\ref{fig: sim_data_points}).
    \item The binary mass is typically significantly greater than the initial total mass within the Hill sphere, suggesting that the self-gravity of the gas may safely be neglected. 
    \item The typical inspiral timescale for an embedded BBH is significantly longer than its orbital period but shorter than its migration timescale. This justifies neglecting the real-time feedback from gas on the semi-major axis of the BBH and its orbit around the central SMBH.
    \item For the SG03 and TQM05 disk models we find that there exists a radius within which the global response of the disk is faster than the orbital evolution of the binary, thus suggesting that global simulations are necessary to capture accurately the long-term binary evolution. This radius ranges from $10^2$ to $10^5$$R_{\rm s}$ for a typical $50M_\odot$ binary, depending on the specific AGN model and parameters. 
    \item There exists a radial disk location within which the radial flow may influence the binary evolution. For the SG03 model this is $\sim 10^2R_{\rm s}$ for a typical AGN-BBH system, whereas for TQM05 this radius extend to $\sim10^6R_{\rm s}$. Further studies on the effect of strong radial flow on the binary evolution will be needed to understand the importance of this effect, and its consequences for the applicability of local models.
\end{itemize}

\vspace{1cm}

The shearing box is an essential tool for modeling binaries embedded in disks. This is because it concentrates computational resources around the binary, allowing for the inclusion of more complex physical processes. Notably, these include the incorporation of magnetohydrodynamics \citep[e.g.][]{Mishra_2024,Joshi_2025} and radiative transfer \citep[e.g.][]{Hirose_2009,Jiang_2013,Jiang_2021}. These elements are expected to play key roles in the evolution of embedded BBHs in AGN disks, yet remain prohibitively demanding to model within fully global simulations. In this work, we have presented a systematic way to frame, and assess the accuracy of, shearing box models of disk-embedded binaries into the global context they seek to model. Having the ability to assess the appropriateness of these models will reinforce the astrophysical relevance of more complex shearing-box studies as we seek to understand AGN disks as promising environments for black hole binary mergers.

\acknowledgments
We thank Chang-Goo Kim for providing us with his first-order flux correction implementation in \texttt{Athena++}. The research leading to this work received funding from the Independent Research Fund Denmark via grant ID 10.46540/3103-00205B. R.L. acknowledges support from the Heising-Simons Foundation 51 Pegasi b Fellowship. The Tycho supercomputer hosted at the SCIENCE HPC center at the University of Copenhagen was used in this work. 

\appendix

\section{How to apply our framework}
\label{sec:apply_framework}
\begin{figure*}[ht!]
\includegraphics[width=\linewidth]{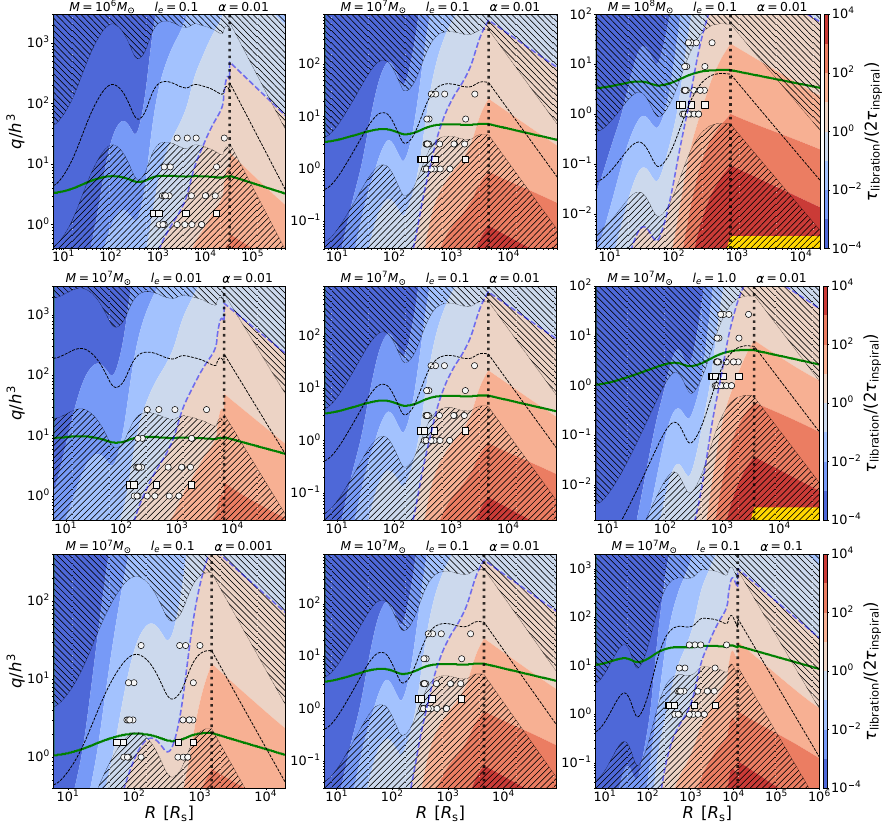}
\caption{Similar to upper right panel of Figure~\ref{fig: sg_tho_single} but for varying SMBH mass $M$, Eddington accretion fraction $l_e$ and $\alpha$-viscosity of the SG03 disk model. In the yellow region the Hill mass is larger than the binary mass, $\Sigma R_H^2>m_b$, see Equation~(\ref{eq. disk_mass_limit}). Here, we have omitted the arrows introduced in Figure~\ref{fig: alpha_single}, which indicate that the data points shown correspond to lower limits.}
\label{fig: sg_3x3}
\end{figure*}

\begin{figure*}[ht!]
\includegraphics[width=\linewidth]{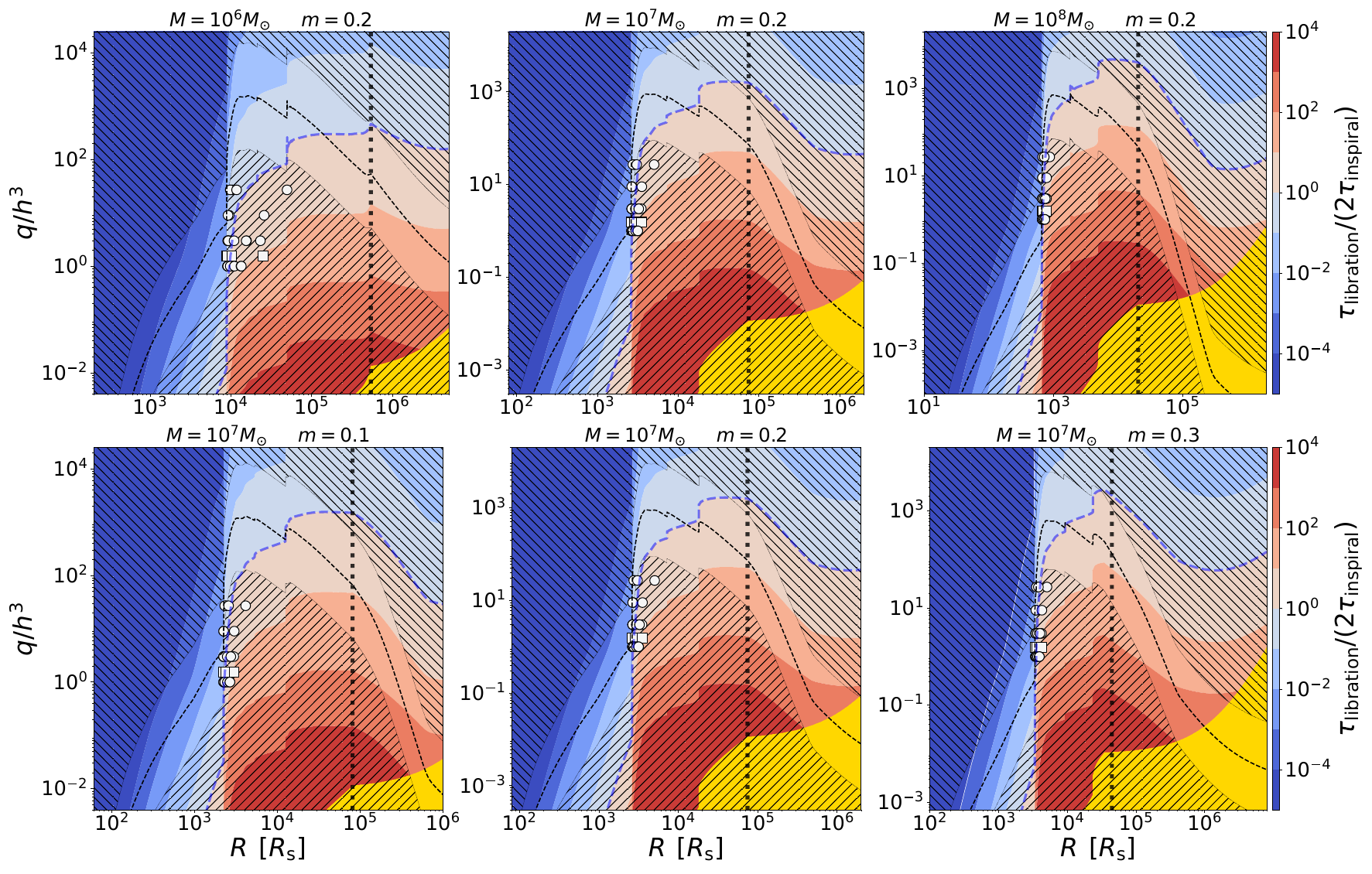}
\caption{Similar to lower right panel of Figure~\ref{fig: sg_tho_single} but for varying SMBH mass $M$ and accretion efficiency $m$ of the TQM05 disk model. 
In the yellow region the Hill mass is larger than the binary mass, $\Sigma R_H^2>m_b$, see Equation~(\ref{eq. disk_mass_limit}). Here, we have omitted the arrows introduced in Figure~\ref{fig: alpha_single}, which indicate that the data points shown correspond to lower limits.}
\label{fig: tho_3x3}
\end{figure*}

We have presented a framework for placing the setups and results of local simulations of embedded black hole binaries into the global context of AGN disks. Figures~\ref{fig: sg_3x3} and \ref{fig: tho_3x3} can be used both to motivate the choice of $q/h^3$ when setting up a simulation and, a posteriori, to assess whether the results are valid for describing the long-term evolution of embedded BBHs in AGN disks. We briefly recapitulate how this is done. 

In each panel of Figures~\ref{fig: sg_3x3} and \ref{fig: tho_3x3}, a chosen value of $q/h^3$ defines a horizontal line. The corresponding binary mass as a function of radius is estimated by comparing this horizontal line to the dotted black curve representing $m_b=50M_{\odot}$, and the right- and left-hashed regions for which $m_b<5M_{\odot}$ and $m_b>500M_{\odot}$, respectively. That is, only radii at which the horizontal line crosses the un-hashed region represents typical BBH systems with $m_b\in[5,500]M_{\odot}$. The ratio of the libration to inspiral timescale is then read off by comparing the horizontal line, defined from the chosen value of $q/h^3$, with the coloured contours. Where the horizontal line intersects red contours, the libration timescale exceeds the inspiral timescale (assuming $\tau'_{\rm inspiral}=1$), indicating radii where local simulations are expected to capture the binary’s long-term evolution. If the value of $\tau'_{\rm inspiral}$ is known, the contours can simply be rescaled by this value, shifting the red–blue division---inwards for $\tau'_{\rm inspiral}<1$ and outwards for $\tau'_{\rm inspiral}>1$.

\bibliographystyle{apj.bst}
\bibliography{bibliography.bib}

\end{document}